\documentclass[twocolumn,showpacs,preprintnumbers,amsmath,amssymb]{revtex4}

\usepackage{graphicx}
\usepackage{dcolumn}
\usepackage{bm}

\begin{document}


\title{Anomalous exciton lifetime by an electromagnetic coupling of self-assembled quantum dots}

\author{E.W. Bogaart}\author{J.E.M. Haverkort}\affiliation{COBRA Inter-University Research Institute, Eindhoven University of Technology, Department of Applied Physics, P.O. Box 513, 5600 MB Eindhoven, The Netherlands}

\begin{abstract}
We report on the experimental observation of a hitherto ignored long-range electromagnetic coupling between self-assembled quantum dots. A 12 times enhancement of the quantum dot exciton lifetime is observed by means of time-resolved differential reflection spectroscopy. The enhancement is explained by utilizing and extending the local field effects as developed in \emph{Phys. Rev. B \textbf{64},125326 (2001)}. The electromagnetic coupling of the quantum dots results in a collective polarizability, and is observed as a suppression of the emission rate. Our results reveal that the coupling is established over a distance exceeding 490 nm. Moreover, the mutual coupling strength is optically tuned by varying the pump excitation density and enables us to optically tune the exciton lifetime.\end{abstract}

\pacs{78.47.+p,42.50.Fx,78.67.Hc}
\maketitle

Realization of nanoscaled semiconductor systems, e.g., quantum dots (QDs), has opened up the possibility of investigating the effects of the local electrodynamic environment on the radiative emission of confined excitons. One of the consequences of a modified electromagnetic density of states (DOS) is the enhancement or suppression of the spontaneous emission rate. Modification of the local density of electromagnetic states in the vicinity of an emitter can be established by inserting the emitter into a nanocavity \cite{vah03} or by the mutual interaction between closely spaced emitters. The latter can be described by F$\ddot{o}$rster coupling \cite{for59,ala04}, the Dicke-effect \cite{dic54,par05}, and local field effects \cite{sle01}. The Dicke-effect is observed as an enhancement or suppression of the emission rate - known as superradiance and subradiance, respectively - and is sensitive to the distance between the interacting emitters. To observe super- or subradiance in solid state nanostructures, the system has to be engineered with high accuracy as established for a high-quality multiple quantum well structure \cite{hub96}. Unlike quantum wells, QD nanostructures have a significant inhomogeneously broadened DOS and a lateral distribution. As a result super- and subradiance tend to cancel each other \cite{par05}.

Experimentally, it is ignored that optical excitation of nanoscale objects, e.g., QDs, quantum rings, and carbon nanotubes, has a profound impact on the permittivity of the nanoscaled object and near its lattice site. For InAs/GaAs QDs this means that due to the Lorenzian-shaped permittivity of optically excited QD $\varepsilon_{QD}(\omega)$, which is much larger than the permittivity of the GaAs host medium $\varepsilon_h$ \cite{sch87,mak00, noteper}, the electromagnetic field pattern is strongly modified. Hereby, the radiative decay channels \cite{sle01,bon02} of the emitter are affected. In addition, the high contrast permittivity landscape formed by excited QDs provides a strong dipole scattering of electromagnetic fields \cite{sle01}, such that local electromagnetic fields acting on each QD are modified by the adjacent polarized QDs. In this perspective, a QD ensemble can be described as an array of mutually interacting point sources, the strength of which is given by the self-dressed polarizability.

In this Letter, we report on a significant modification of the QD-exciton lifetime observed by means of time-resolved differential reflection spectroscopy (TRDR) \cite{bog05a}, owing to mutual electromagnetic coupling between QD-excitons within the ensemble \cite{bog06c,bog07}. The profound modification of the QD-exciton lifetime is a direct consequence of the collective polarizability of the nanostructure. To explain this collective effect, we utilize and significantly extend the electromagnetic response theory as presented by Ref. \cite{sle01} for a two-dimensional array of identical QDs. Our extended model describes the effective electromagnetic response of nanostructures, which takes into account the inhomogeneously broadened DOS as observed for more realistic QD ensembles. The significant lifetime enhancement illustrates a mutual interaction between resonant QD-excitons, which extends over distances considerably beyond the nearest neighbor separation.

By plotting the QD-exciton lifetime versus the QD transition energy an anomalous QD-exciton lifetime spectrum is revealed, as depicted in Fig. 1. The lifetime spectrum - deduced from our transient differential reflection measurements - has a pronounced resonant-like behavior with respect to the optically excited QD DOS with a 12 times enhancement at the mean transition energy of the QD ground state.

Let a single random inhomogeneous layer of QDs be placed in a homogeneous isotropic host medium. Our model incorporates two factors of the irregularity: i) inhomogeneous broadening of the exciton spectrum, and ii) random distribution of QDs in the layer. As a first step, we assume all QDs to be identical and eliminate thus the inhomogeneous broadening. Then, the effective boundary conditions derived for regular arrays in Ref. \cite{sle01}, are extended to the case of randomly inhomogeneous 2D ensembles of QDs with homogeneous line broadening. The electromagnetic response utilizes the Clausius-Mossotti relation \cite{asp82,jac75} to calculate the collective polarizability of the ensemble. Further analysis is based on general principles of the wave propagation in randomly inhomogeneous media.

The resulting local field effects are visualized by the optical response functions. We use the reflection coefficient - s-polarization - of a planar periodic array of identical QDs with an effective volume $V_{QD}$ to determine the collective radiative decay rate. The emission rate of an electromagnetically coupled QD array $\Gamma^{coupled}$  - at resonance $\hbar\omega_0$ - is governed by the imaginary part of the frequency pole of the reflection coefficient \cite{sle01}, and can be written as
\begin{equation}
\Gamma^{coupled} = \Gamma^{isolated} -\frac{k_0 \sqrt{\varepsilon_h}\omega_{QD}V_{QD}}{2d^2_{QD}}.
\end{equation}
Here $\Gamma^{isolated}$ is a weighted sum of the dephasing and the radiative emission rate of an isolated QD-exciton, i.e., an uncoupled QD-exciton \cite{sug95,bel98,bim99,bor01,lan04}. $k_0$ and $d_{QD}$ denote the vacuum wave vector and the spacing of the QDs within the two-dimensional periodic lattice, respectively. $\omega_{QD}$ is the phenomenological parameter proportional to the QD oscillator strength \cite{sle01}. Equation (1) shows that the lifetime 1/$\Gamma^{coupled}$ is governed by the lattice spacing $d_{QD}$.

To apply our theoretical model to self-assembled QD nanostructures, the inhomogeneously broadened DOS and random spatial QD distribution must be introduced in the model. We emphasize, that the electromagnetic coupling manifests itself only between (quasi)resonant QD-excitons and is governed by the spectral overlap of the homogeneously broadened permittivity of the excited QDs, typically described by a Lorenzian profile. Lateral disorder means distortion of the ideal lattice, and is seen as a fluctuation of distances between adjacent sites. Disorder effects of excitons in semiconductor nanostructures can be described within the usual single-site approximation, e.g., the coherent-potential approximation (CPA) \cite{ste93,mod00,per86}. If $Z$ different scatterers are randomly distributed on the lattice sites of a plane, the CPA can be used to treat the disordered system within a mean-field context. Hereby, the effective CPA medium consists of identical scatterers at all sites of the plane, i.e., an isotropic QD-array. By making use of the CPA formulism, the whole QD ensemble with inhomogeneously broadened DOS $G(\omega)$ can be divided into smaller sub-ensembles, each with their own narrow DOS $G(\omega_j)$ and unique ordering - a two-dimensional array with periodicity $d^{res}_{QD}(\omega_j)$, where each lattice site is occupied by a QD-exciton with the effective polarizability.

The average distance between the QDs within each sub-ensemble depends on the location of these QDs within the overall DOS, 1/$d^{res}_{QD}(\omega_j)$ = $\sqrt{N(\omega_j)}$ = $\sqrt{N_{QD}}G(\omega_j)$. Here $N(\omega_j)$ and $N_{QD}$ are the area QD density of a sub-ensemble with peak energy $\hbar\omega_j$ and of the whole ensemble, respectively. Taking this into account for the whole ensemble, Eq. (1) is rewritten as
\begin{equation}
\Gamma^{coupled}(\omega) = \Gamma^{isolated} -2\frac{k_0 \sqrt{\varepsilon_h}\omega_{QD}V_{QD}}{2}N_{QD}G^2(\omega).
\end{equation}
The additional factor two in the second term on the righthand side takes into account the QD spin degeneracy. Equation (2) reveals that the QD-exciton lifetime has a maximum at the center of the QD-size distribution, where the average distance between radiatively coupled QDs within the sub-ensemble has statistically a minimum value. Equation (2) also predicts a resonant-like QD-exciton lifetime spectrum. The lifetime spectrum will have a narrower spectral width - a factor $\sqrt{2}$ narrower - than the QD distribution $G(\omega)$, due to the quadratic dependence. This prediction is indeed observed experimentally, as will be shown below.

To test the result of the theoretical electromagnetic response model as given by Eq. (2), differential reflection measurements are performed on a self-assembled InAs QD nanostructure grown by molecular beam epitaxy on a (100) GaAs substrate. The nanostructure studied here, is the same sample as is reported in Ref. \cite{bog05a} and has a density of 2.8$\times$10$^{10}$ QDs cm$^{-2}$. The QD-exciton lifetime is investigated by means of pump-probe TRDR \cite{bog05a,oth98,bog06a} at 5 K. The QDs are non-resonantly excited using short laser pulses from a mode-locked Ti:sapphire laser. The pump photon energy is tuned above the bandgap energy of the GaAs barrier material and creates free carriers within the GaAs host, which are subsequently captured into the QDs. The carrier-induced reflection change $\Delta R$/$R_0 (\omega)$ is monitored by tuning the photon energy of the probe laser over the QD optical transition energy within the ensemble. These resonant probe pulses are generated from an optical parametric oscillator, synchronically pumped by the Ti:sapphire laser. Both lasers emit 2 ps pulses, corresponding to a spectral resolution of approximately 1 meV. Thus, in the center of the size distribution, only 1.7$\%$ of all QDs are in resonance with the probe laser. This small fraction of resonant QDs forms an interactive coupled sub-ensemble, which collectively responds to the electromagnetic probe field. Since the average dot-to-dot distance in our sample is approximately 60 nm, the average separation between resonant dots extends to approximately 460 nm in the center of the size distribution.

Figure 1 depicts the QD-exciton lifetime as a function of the QD transition energy. We note that the QD-exciton lifetime spectrum shows a pronounced resonant-like behavior with respect to the photoluminescence (PL) spectrum. The lifetime spectrum reveals, that the QD-exciton lifetime at the mean transition energy of the QD ground state is 12 times larger than that of the excitons at the wings of the distribution. In other words, Fig. 1 reveals a 12 times enhancement of the QD-exciton lifetime. Moreover, the lifetime spectrum is narrower than the PL spectrum which is in agreement with the predictions as deduced from Eq. (2). Surprisingly, also the lifetime of the QD first excited state at 1.163 eV has an energy dependence. These preliminary observations already provide some qualitative experimental confirmations of our theoretical model. However to make a more quantitative statement, additional investigations of the TRDR measurement results have to be performed. Analysis of the lifetime spectrum by using Gaussian fits, reveals two peaks which are ascribed to the QD ground and first excited state with peak energies of 1.104 and 1.161 eV, respectively. In addition, the spectral width of the ground (first excited) state is 27 meV (23 meV). The PL spectrum has a peak energy of 1.107 eV for the ground state and a spectral width of 44 meV. We note, that the lifetime spectrum is slightly shifted with respect to the PL spectrum.

At the side of the distribution the QD-exciton lifetime approaches 150 ps. These QDs can be regarded as uncoupled and enable us to determine $\Gamma^{isolated}$. Our observation of a 150 ps QD-exciton lifetime is not surprising if we consider that our QDs have a total confinement energy of approximately 168 meV \cite{bog05a}. From the results as reported in Ref. \cite{lan04}, a radiatively limited lifetime of a few hundred picoseconds is expected.

More experimental evidence of electromagnetic coupling between optically excited QDs is provided by the pump excitation density dependence of the QD-exciton lifetime enhancement, as depicted in Fig. 2. From the spectra shown in Figs. 1 and 2, it is clear that the lifetime-spectrum amplitude exhibits a strong dependence on the pump-induced carrier density $\eta$, but that the shape remains unaffected. The increased pump-induced carrier density induces a higher QD population, resulting in a nonzero change of the QD-exciton emission rate \cite{bog05a,bog06a}, $\Delta \Gamma$ = $\partial \Gamma / \partial \eta$ $\neq$ 0. Hence, the QD-exciton lifetime is governed by the density of occupied states, $\Gamma^{coupled} (\omega)\sim G^2_{occupied}(\omega)$. Thus, the number of optically excited QDs determines the effective coupling distance and hereby the coupling strength.

To compare our theoretical model with the experimental findings, Figs. 1 and 2, the QD DOS $G(\omega)$ has to be known. We utilize the differential reflection spectrum obtained by plotting the amplitude of the TRDR time-traces, as depicted in the inset of Fig. 1, versus the probe photon energy. The resultant reflectivity spectrum is depicted in Fig. 3. Two peaks are observed for the QD ground state and for the QD first excited state. Both energy states have a local minimum near the peak of the ground and first excited state transition energy. I.e., for both energy states a splitting is observed in the differential reflection spectrum and appears as a double peak shifted $\pm \hbar\omega_c$ with respect to the mean transition energy of each state. From the reflectivity spectrum, we deduce a splitting of 2$\hbar\omega_c$ = 27 (15) meV for the QD ground (first excited) state.

The amplitude $\Delta R(\omega)$ of the differential reflectivity, Fig. 3, is expressed as \cite{bog05a,bog06a}
\begin{eqnarray}
\Delta R(\omega) &=& -\int H(\omega -\omega')G(\omega')d\omega',\nonumber \\
\textrm{   with   }& & H(\omega) = \Delta\Gamma(\omega)[\Gamma(\omega)L'(\omega)+L(\omega)],
\end{eqnarray}
in which $L(\omega)$ is a Lorentzian line shape function of a single isolated QD modelling the homogeneous broadening, and $L'(\omega)$ is its first derivative \cite{bog05a,bog06a}. Combining the measured reflectivity spectrum $\Delta R$/$R_0(\omega)$ with the pump excitation dependence of the emission rate $\Delta\Gamma (\omega)$, as deduced from Figs. 1 and 2, the energy distribution $G(\omega)$ is extracted by using Eq. (3). The result is depicted in Fig. 4.

From Eq. (2) it is expected, that the width of the lifetime spectrum is a factor $\sqrt{2}$ narrower than the DOS. This means, we have to compare the lifetime spectra with the extracted distribution $G(\omega)$ and not with the PL spectrum. From our analysis we observe, that the extracted distribution is described by a superposition of two Gaussian functions with mean energies of 1.103 and 1.154 eV, respectively. The main peak, ascribed to the QD ground state, has a spectral width of approximately 38 meV. The theoretical prediction is accurately verified since the 27 meV spectral width of $\Gamma^{coupled}(\omega)$ is found to be a factor 1.41 smaller than the 38 meV spectral width of $G(\omega)$. Thus, our theoretical model of electromagneticly coupled QD-excitons is experimentally verified.

Now let us determine parameter $\omega_{QD}$ from which we can derive the mutual coupling distance $d^{res}_{QD}$. We employ the experimentally observed splitting in the TRDR spectrum (Fig. 3). This splitting can be described in terms of an exciton-photon coupling - inducing a QD-polariton splitting - with energy $\hbar\omega_c$, and is a direct measure of the mixing between the QD-exciton and photon states. This allows us to derive the longitudinal-transverse splitting $\hbar\omega_{LT}$ in QDs \cite{ivc00,and95}, $\omega_c$ = $\sqrt{\frac{\omega_0\omega_{LT}}{2}}$. From the polariton splitting of the ground state transition, 2$\hbar\omega_c$ = 27 meV, an effective LT-splitting of 0.33 meV is determined. Finally, we obtain the phenomenological parameter $\omega_{QD}$ = 3$\omega_c$. We emphasize, the ratio $\omega_{LT}$/$\omega_0$ $\approx$ 3$\times 10^{-4}$ is in perfect agreement with the value typically observed for semiconductors \cite{ivc00}.

Applying Eq. (2) to our experimental results, the 12 times enhancement of the exciton lifetime at the center of the QD-size distribution - deduced from the data depicted in Fig. 1 - corresponds to an average distance of $d^{res}_{QD}$ = 490 $\pm$ 20 nm between the radiatively coupled QD-excitons. We note, that the exciton lifetime at the wings of the distribution are thus governed by an even larger separation of the mutually interacting QDs. These QDs have a weaker coupling strength and therefore a smaller lifetime enhancement.

Our observation of a long-range electromagnetic interaction between QD-excitons, implies that the photonic lattices formed by QDs \cite{ivc00} are promising systems for the development of a quantum processor \cite{los98}. Because long-range interaction mechanisms in which the interaction is not limited to the nearest neighbors, are essential for building a scalable quantum computer \cite{pet02}. We expect that our results will open intriguing perspectives for the emerging fields of quantum logics, in which the photonic lattice can be regarded as a quantum register with each QD-exciton on a lattice site acting as a qubit.

In summary, electromagnetic interaction between distant QDs is observed from transient differential reflectivity measurements. The QD-exciton lifetime is measured as a function of the transition energy and shows a strong resonant-like behavior with respect to the QD DOS. The obtained lifetime spectrum reveals a 12 times lifetime enhancement at the center of the ground state energy distribution, due to the collective effect of electromagneticly coupled QDs.

The authors thank R. N\"{o}tzel, G.Ya. Slepyan, S.A. Maksimenko, V. Savona, and D.A.M. Vanmaekelbergh for fruitful discussions. This work is financially supported by the Stichting voor Fundamenteel Onderzoek der Materie (FOM).
\newpage
\bibliography{biblio}

\begin{thebibliography}{30}
\bibitem[{not()}]{noteper}
\bibinfo{note}{From Ref. \cite{mak00}, an excited QD has a Lorenzian-shaped
  permittivity with a calculated peak value of $\varepsilon_{QD,peak} \sim
  10^6$ at resonance.}
\end{thebibliography}
\newpage

\begin{figure}
\includegraphics[width=0.8\linewidth]{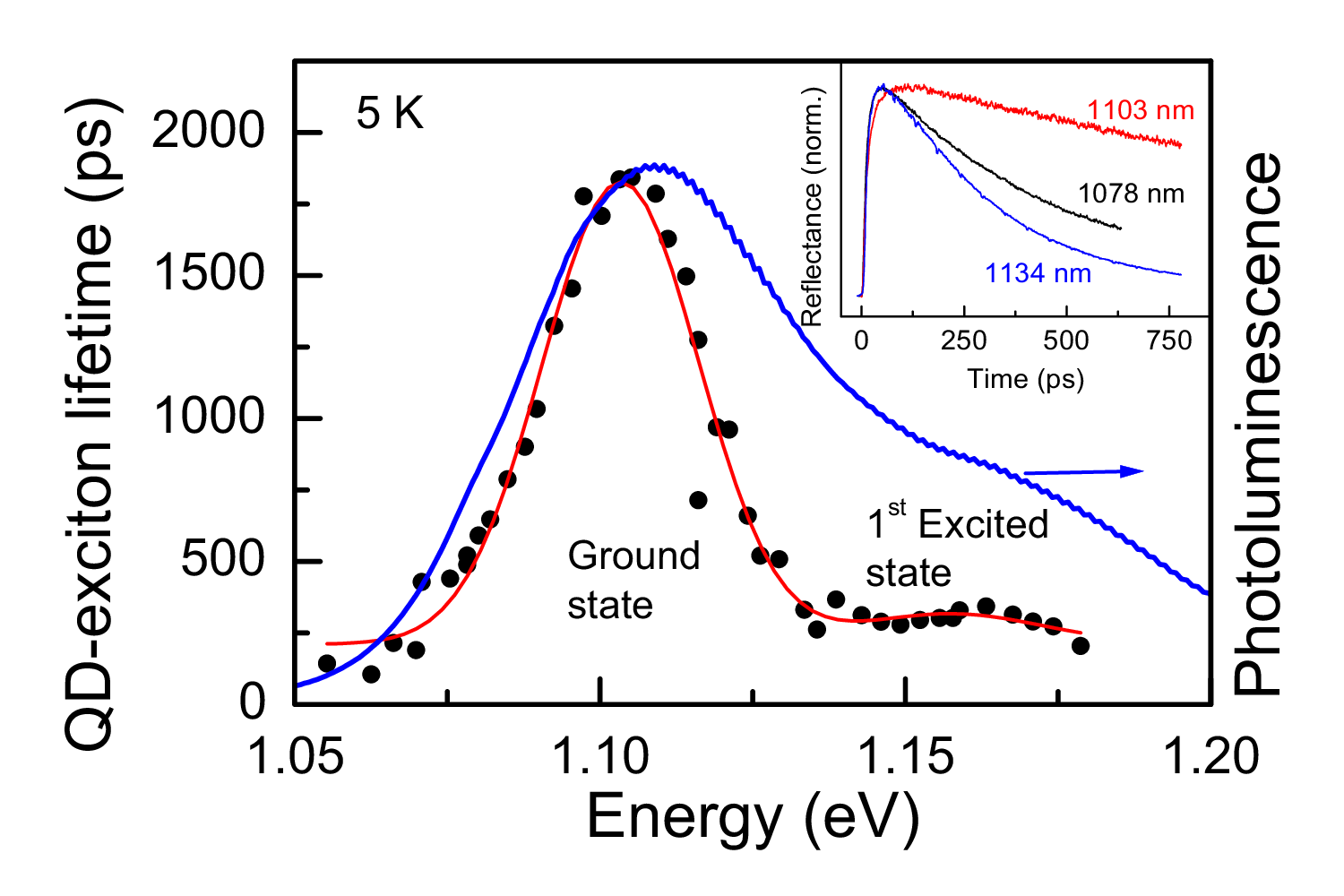}
\caption{\label{fig1}(Color online) QD-exciton lifetime versus the probe photon energy measured at 300 W/cm$^2$ pump excitation, revealing anomalous lifetime resonance for the QD ground and excited state. A PL spectrum obtained at high excitation density is added as a reference, to emphasize the QD first excited state. The inset shows three TRDR signals (normalized) obtained for different probe photon energies.}
\end{figure}
\begin{figure}
\includegraphics[width=0.8\linewidth]{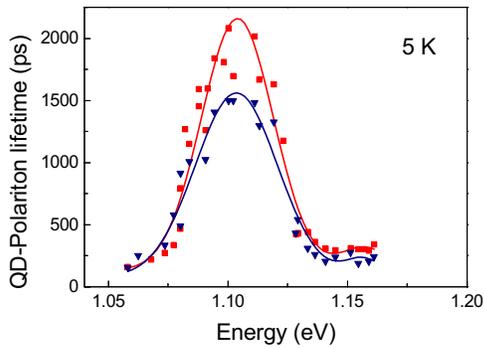}
\caption{\label{fig2}(Color online) QD-exciton lifetime versus the probe photon energy obtained for a pump excitation density of 200 (triangle) and 400 W/cm$^2$ (square), corresponding to approximately one and two electron-hole pairs in a single QD, respectively.}
\end{figure}
\begin{figure}
\includegraphics[width=0.8\linewidth]{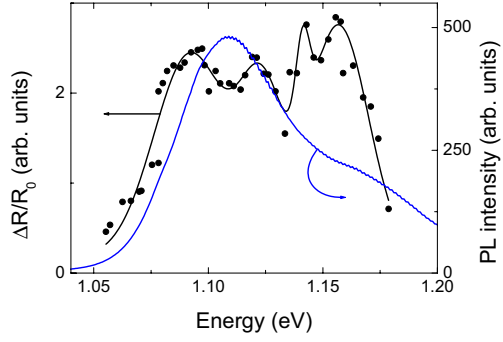}
\caption{\label{fig3}(Color online) Differential reflection spectrum and the QD PL spectrum, both measured at 5 K. The line through the reflection spectrum is only a guide for the eye.}
\end{figure}
\begin{figure}
\includegraphics[width=0.8\linewidth]{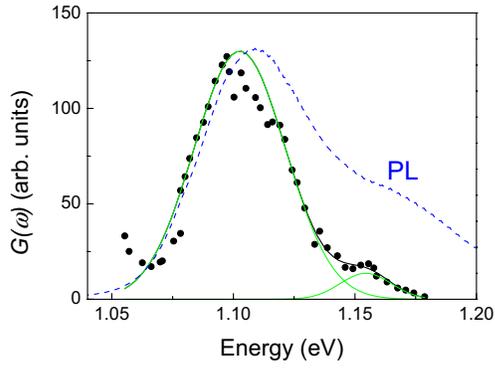}
\caption{\label{fig4}(Color online) QD DOS $G(\omega)$ extracted from the differential reflection spectrum as depicted in Fig. 3. The full curves illustrate the Gaussian fit to the size distribution, in order to highlight the realistic nature of the extracted $G(\omega)$. The PL spectrum (dashed) is provided as a reference.}
\end{figure}

\end{document}